\documentclass[11pt,titlepage,a4paper]{article}
\usepackage{bozhomac}	
\usepackage{bozhlogo}	
\usepackage{cite}	

\title{\bfseries	\vspace*{-1.678902345in}
\vspace*{-7ex}
{
\begin{flushright}
\textbf{\large LANL arXiv server, E-print No. hep-th/0110194}\\[5ex]
\end{flushright}
}
{\huge Normal frames for linear\\ connections in vector bundles\\
	and the equivalence principle\\ in classical gauge theories
}
}

\vspace{1.7ex}

\author{
Bozhidar Z. Iliev
\thanks{Department Mathematical Modeling,
Institute for Nuclear Research and \mbox{Nuclear} Energy,
Bulgarian Academy of Sciences,
Boul. Tzarigradsko chauss\'ee~72, 1784 Sofia, Bulgaria}
\thanks{E-mail address: bozho@inrne.bas.bg}
\thanks{URL: http://theo.inrne.bas.bg/$\sim$bozho/}
}

\date{
 \vspace{2.27ex}\ShortTitle{Normal frames, connections, and gauge fields}
								\\[0.27ex]
 \vspace{3.27ex}
	\begin{tabular}{r@{$\colon\to~$}l}
 \vspace{0.09ex} Basic ideas	& November 1998--October 2000	\\[0.09ex]
 \vspace{0.09ex} Began		& October 25, 2000	\\[0.09ex]
 \vspace{0.09ex} Ended		& November 17, 2000	\\[0.09ex]
\vspace{0.09ex} Initial typeset& November 20, 2000 -- December 13, 2000
							\\[0.09ex]
 \vspace{0.09ex} Last update	& October 21, 2001	\\[0.09ex]
 \vspace{0.27ex} Produced	& \fbox{\today}	\\[0.27ex]
	\end{tabular} \\[1.27ex]
	\begin{tabular}{r@{$\colon~$}l}
\vspace{0.27ex} LANL xxx archive server E-print No. & hep-th/0110194
						\\[0.27ex]
	\end{tabular} \\[-0.27ex]
 \vspace{4.27ex}{\Huge\BOZHO}	\\[4.27ex]
 \vspace{0.27ex}\Subject{Differential geometry, Gauge
			 theories, Unified field theories}	\\[2.27ex]
	\begin{tabular}{r@{\hspace{0.512em}}|@{\hspace{0.512em}}l}
 \vspace{0.27ex}	
&
 \vspace{0.27ex} 
	\end{tabular} \\[1.27ex]
 \vspace{0.27ex}	\\[0.27ex]
}

\newcommand{\Esf}{\mathsf{E}}

\begin{document}		

\renewcommand{\thefootnote}{\fnsymbol{footnote}} 
\maketitle				
\renewcommand{\thefootnote}{\arabic{footnote}}   

\tableofcontents		


\begin{abstract}

	Frames normal for linear connections in vector bundles are defined
and studied. In particular, such frames exist at every fixed point and/or
along injective path. Inertial frames for gauge fields are introduced and on
this ground the principle of equivalence for (system of) gauge fields is
formulated.

\end{abstract}

\section {Introduction}
\label{Introduction}

	Until 1992 the existence of normal frames and coordinates was known
at a single point and along injective paths only for symmetric linear
connections on a manifold. The
papers~\cite{bp-Frames-n+point,bp-Frames-path,bp-Frames-general} (see also
their early versions~\cite{bp-Bases-n+point,bp-Bases-path,bp-Bases-general})
completely solved the problems of existence, uniqueness and holonomicity of
frames normal on submanifolds for derivations of the tensor algebra over a
manifold, in particular for arbitrary, with or without torsion, linear
connections on a manifold. At last, these results were generalized
in~\cite{bp-normalF-LTP} for linear transports along paths in vector bundles.
The present work can be considered as a continuation as well as an
application of the cited references. Here we investigate problems concerning
frames normal for arbitrary linear connections in vector bundles and show
that the already existing results can \emph{mutatis mutandis} be applied in
this situation.

	Sect.~\ref{Sect2} recalls the most suitable for us definition of a
linear connection in a vector bundle and some consequences of it.
Sect.~\ref{Sect3} summarizes basic concepts of the theory of linear
transports along paths in vector bundles. In Sect.~\ref{Sect4} are proved
necessary and sufficient conditions for a derivation or a linear transport
along paths in vector bundles to defined a linear connection. An explicit
bijective correspondence between a particular class of such objects and the
set of linear connections is derived. The parallel transports generated by
linear connections are described in terms of linear transports along paths.

	In Sect~\ref{Sect5} the frames normal for linear connections in
vector bundles are defined and the basic equation responsible for their
existence and properties is derived. Since this equation coincides with
similar equations investigated
in~\cite{bp-Frames-n+point,bp-Frames-path,bp-Frames-general}, the conclusion
is made that the results of these papers can \emph{mutatis mutandis} be
applied to solved similar problems concerning frames normal for linear
connections in vector bundles. Some particular results are written
explicitly.

	In Sect~\ref{Sect6} is shown how inertial frames in gauge field
theories should be introduced. The principle of equivalence, which in fact is
a theorem, for a particular gauge field is formulated. An example is
presented for the introduction of inertial frames and formulation of the
equivalence principle for a system of gauge fields (and, possibly,
gravitational one).

	Sect~\ref{Conclusion} ends the work.


\section {Linear connections in vector bundles}
\label{Sect2}

	Different equivalent definitions of a (linear) connection in vector
bundles are known and in current usage~\cite{K&N,Poor, Greub&et_al.,
Bleecker}. The most suitable one for our purposes is given
in~\cite[p.~223]{Baez&Muniain} (see also~\cite[theorem~2.52]{Poor}).

	Suppose $(E,\pi,M)$, $E$ and $M$ being finite-dimensional  $C^\infty$
manifolds, be $C^\infty$ \field-vector bundle~\cite{Poor} with bundle space
$E$, base $M$, and projection $\pi\colon E\to M$. Here \field\ stands for the
field \field[R] of real numbers or \field[C] of complex ones. Let
$\Sec^k(E,\pi,M)$, $k=0,1,2,\dots$ be the set (in fact the module) of $C^k$
sections of $(E,\pi,M)$ and $\Fields{X}(M)$ the one of vector fields on $M$.

	\begin{Defn}	\label{Defn2.1}
	Let $V,W\in\Fields{X}(M)$, $\sigma,\tau\in\Sec^1(E,\pi,M)$, and
$f\colon M\to\field$ be a $C^\infty$ function. A mapping
 $\nabla\colon\Fields{X}(M)\times\Sec^1(E,\pi,M)\to\Sec^0(E,\pi,M)$,
 $\nabla\colon(V,\sigma)\mapsto\nabla_V\sigma$, is called a (linear)
connection in $(E,\pi,M)$ if:
	\begin{subequations}		\label{2.1}
	\begin{gather}	\label{2.1a}
\nabla_{V+W}\sigma = \nabla_{V}\sigma + \nabla_{W}\sigma ,
\\			\label{2.1b}
\nabla_{fV}\sigma = f \nabla_{V}\sigma ,
\\			\label{2.1c}
\nabla_{V}(\sigma+\tau) = \nabla_{V}(\sigma) + \nabla_{V}(\tau) ,
\\			\label{2.1d}
\nabla_{V}(f\sigma) = V(f)\cdot\sigma + f\cdot \nabla_{V}(\sigma) .
	\end{gather}
	\end{subequations}
	\end{Defn}

	\begin{Rem}	\label{Rem2.1}
	Rigorously speaking, $\nabla$, as defined by
definition~\ref{Defn2.1}, is a covariant derivative operator in $(E,\pi,M)$
--- see~\cite[definition~2.51]{Poor} --- but, as a consequence
of~\cite[theorem~2.52]{Poor}, this cannot lead to some ambiguities.
	\end{Rem}

	\begin{Rem}	\label{Rem2.2}
	Since $V(a)=0$ for every $a\in\field$ (considered as a constant
function $M\to\{a\}$), the mapping
$\nabla\colon(V,\sigma)\mapsto\nabla_V\sigma$ is \field-linear with respect
to both its arguments.
	\end{Rem}

	Let $\{e_i: i=1,\dots,\dim\pi^{-1}(x)\}$, $x\in M$ and
$\{E_\mu:\mu=1,\dots,\dim M\}$ be frames over an open set  $U\subseteq M$
in, respectively, $(E,\pi,M)$ and the tangent bundle $(T(M),\pi_T,M)$ over
$M$, \ie for every $x\in U$, the set $\{e_i|_x\}$ forms a basis of the fibre
$\pi^{-1}(x)$ and $\{E_\mu|_x\}$ is a basis of the space
$T_x(M)=\pi_T^{-1}(x)$ tangent to $M$ at $x$. Let us write
$\sigma=\sigma^ie_i$ and $V=V^\mu E_\mu$, where here and henceforth the Latin
(resp.\ Greek) indices run from 1 to the dimension of $(E,\pi,M)$
(resp.\ $M$), the Einstein summation convention is assumed, and
$\sigma^i,V^\mu\colon U\to\field$ are some $C^1$ functions. Then, from
definition~\ref{Defn2.1}, one gets
	\begin{equation}	\label{2.2}
\nabla_V\sigma =
  V^\mu\bigl(
	E_\mu(\sigma^i) + \Sprindex[\Gamma]{j\mu}{i} \sigma^j
  \bigr) e_i
	\end{equation}
where $\Sprindex[\Gamma]{j\mu}{i}\colon U\to\field$, called
\emph{coefficients} of $\nabla$, are given by
	\begin{equation}	\label{2.3}
\nabla_{E_\mu}e_j =: \Sprindex[\Gamma]{j\mu}{i} e_i .
	\end{equation}

	Evidently, due to equation~\eref{2.2}, the knowledge of
$\{\Sprindex[\Gamma]{j\mu}{i}\}$ in a pair of frames $(\{e_i\},\{E_\mu\})$
over $U$ is equivalent to the one of $\nabla$ as any transformation
\(
(\{e_i\},\{E_\mu\})
  \mapsto
	(\{e'_i=A_i^je_j\} , \{E'_\mu=B_\mu^\nu E_\nu\})
\)
with non\ndash degenerate matrix\ndash valued functions $A=[A_i^j]$ and
$B=[B_\mu^\nu]$ on  $U$ implies
\(
\Sprindex[\Gamma]{j\mu}{i}
  \mapsto
	\Sprindex[\Gamma]{j\mu}{\prime\mspace{0.7mu} i}
\)
with
	\begin{multline}	\label{2.4}
\Sprindex[\Gamma]{j\mu}{\prime\mspace{0.7mu} i}
  = \sum_{\nu=1}^{\dim M} \sum_{k,l=1}^{\dim\pi^{-1}(x)}
	B_{\mu}^{\nu} \bigl(A^{-1}\bigr)_{k}^{i} A_{j}^{l}
		\Sprindex[\Gamma]{l\nu}{k}
\\
	+ \sum_{\nu=1}^{\dim M} \sum_{k=1}^{\dim\pi^{-1}(x)}
		B_{\mu}^{\nu}\bigl(A^{-1}\bigr)_{k}^{i}
		E_\nu (A_{j}^{k})  .
	\end{multline}
which in a matrix form reads
	\begin{equation}	\label{2.5}
\Mat{\Gamma}_\mu^{\prime}
  = B_{\mu}^{\nu} A^{-1}\Mat{\Gamma}_\nu A  + A^{-1} E'_\mu(A)
  = B_{\mu}^{\nu} A^{-1}\bigl( \Mat{\Gamma}_\nu A  + E_\nu(A) \bigr)
	\end{equation}
where
$\Gamma_\mu:=[\Sprindex[\Gamma]{j\mu}{i}]_{i,j=1}^{\dim\pi^{-1}(x)}$,
$x\in M$, and
\(
\Gamma'_\mu
	:=[\Sprindex[\Gamma]{j\mu}{\prime\,i}]_{i,j=1}^{\dim\pi^{-1}(x)} .
\)

	The interpretation of the coefficients $\Sprindex[\Gamma]{j\mu}{i}$
as components of a 1\ndash form (more precisely, of endomorphisms of $E$
-valued 1\ndash form or section of the endomorphism bundle of $(E,\pi,M)$, or
of Lie algebra\ndash valued 1\ndash form in a case of principle bundle) is
well known and considered at length in the
literature~\cite{Poor,Bleecker,Baez&Muniain,K&N-1,Drechsler&Mayer} but it
will not be needed directly in the present work.


\section {Linear transports along paths in vector bundles}
\label{Sect3}

	To begin with, we recall some definitions and results
from~\cite{bp-normalF-LTP}.%
\footnote{%
In~\cite{bp-normalF-LTP} is assumed $\field=\field[C]$ but this choice is
insignificant.%
}
Below we denote by $\PLift^k(E,\pi,M)$ the set of liftings of $C^k$ paths
from $M$ to $E$ such that the lifted paths are of class $C^k$,
$k=0,1,\ldots$. Let $\gamma\colon J\to M$, $J$ being real interval, be a path
in M.

	\begin{Defn}	\label{Defn3.1}
	A linear transport along paths in vector bundle $(E,\pi,B)$ is a
mapping $L$ assigning to every path $\gamma$ a mapping $L^\gamma$, transport
along $\gamma$, such that $L^\gamma\colon (s,t)\mapsto L^\gamma_{s\to t}$
where the mapping
	\begin{equation}	\label{3.1}
L^\gamma_{s\to t} \colon  \pi^{-1}(\gamma(s)) \to \pi^{-1}(\gamma(t))
	\qquad s,t\in J,
	\end{equation}
called transport along $\gamma$ from $s$ to $t$, has the properties:
	\begin{alignat}{2}	\label{3.2}
L^\gamma_{s\to t}\circ L^\gamma_{r\to s} &=
			L^\gamma_{r\to t},&\qquad  r,s,t&\in J, \\
L^\gamma_{s\to s} &= \id_{\pi^{-1}(\gamma(s))}, & s&\in J,	\label{3.3}
\\
L^\gamma_{s\to t}(\lambda u + \mu v) 				\label{3.4}
  &= \lambda L^\gamma_{s\to t}u + \mu L^\gamma_{s\to t}v,
	& \lambda,\mu &\in \field,\quad u,v\in{\pi^{-1}(\gamma(s))},
	\end{alignat}
where  $\circ$ denotes composition of maps and
$\id_X$ is the identity map of a set $X$.
	\end{Defn}

	\begin{Defn}	\label{Defn3.2}
	A derivation along paths in $(E,\pi,B)$ or a derivation of
liftings of paths in $(E,\pi,B)$ is a mapping
	\begin{subequations}	\label{3.5}
	\begin{equation}	\label{3.5a}
	D\colon\PLift^1(E,\pi,B) \to \PLift^0(E,\pi,B)
	\end{equation}
	\end{subequations}
which is \field-linear,
	\begin{subequations}	\label{3.6}
	\begin{equation}	\label{3.6a}
D(a\lambda+b\mu) = aD(\lambda) + bD(\mu)
	\end{equation}
	\end{subequations}
for $a,b\in\field$ and $\lambda,\mu\in\PLift^1(E,\pi,B)$,
and the mapping
	\begin{equation}
	\tag{\ref{3.5}b}	\label{3.5b}
D_{s}^{\gamma}\colon \PLift^1(E,\pi,B) \to \pi^{-1}(\gamma(s)),
	\end{equation}
defined via
\(
D_{s}^{\gamma}(\lambda)
 := \bigl( (D(\lambda))(\gamma) \bigr) (s)
  = (D\lambda)_\gamma(s)
\)
and called derivation along $\gamma\colon J\to B$ at $s\in J$, satisfies the
`Leibnitz rule':
	\begin{equation}
	\tag{\ref{3.6}b}	\label{3.6b}
D_s^\gamma(f\lambda)
 = \frac{\od f_\gamma(s)}{\od s} \lambda_\gamma(s)
	+ f_\gamma(s) D_s^\gamma(\lambda)
	\end{equation}
for every
\[
f\in
\PF^1(B) :=
	\{
	\varphi|\varphi\colon\gamma\mapsto\varphi_\gamma, \
\gamma\colon J\to B,\ \varphi_\gamma\colon J\to \field
	\text{ being of class $C^1$}
	\} .
\]
The mapping
	\begin{equation}
	\tag{\ref{3.5}c}	\label{3.5c}
D^{\gamma}\colon \PLift^1(E,\pi,B) \to
	\Path\bigl(\pi^{-1}(\gamma(J))\bigr)
	:=\{ \text{paths in } \pi^{-1}(\gamma(J)) \} ,
	\end{equation}
defined by $D^\gamma(\lambda):=(D(\lambda))|_\gamma=(D\lambda)_\gamma$, is
called derivation along $\gamma$.
	\end{Defn}

	If $\gamma\colon J\to M$ is a path in $M$ and $\{e_i(s;\gamma)\}$ is
a basis in $\pi^{-1}(\gamma(s))$,%
\footnote{%
If there are $s_1,s_2\in J$ such that $\gamma(s_1)=\gamma(s_2):=y$, the
vectors $e_i(s_1;\gamma)$ and $e_i(s_2;\gamma)$ need not to coincide. So, if
this is the case, the bases $\{e_i(s_1;\gamma)\}$ and $\{e_i(s_2;\gamma)\}$
in $\pi^{-1}(y)$ may turn to be different.%
}
in the frame $\{e_i\}$ over $\gamma(J)$ the
\emph{components (matrix elements)} $\Sprindex[L]{j}{i}\colon U\to\field$ of
a linear transport $L$ along paths and the ones of a derivation $D$ along
paths in vector bundle $(E,\pi,M)$ are defined through, respectively,
	\begin{gather}	\label{3.7}
L_{s\to t}^{\gamma} \bigl(e_i(s;\gamma)\bigr)
		=:\Sprindex[L]{i}{j}(t,s;\gamma) e_j(t;\gamma)
		\qquad s,t\in J,
\\			\label{3.8}
D_s^\gamma \hat{e}_j =: \Sprindex[\Gamma]{j}{i}(s;\gamma) e_i(s;\gamma),
	\end{gather}
where $s,t\in J$ and $\hat{e}_i\colon\gamma\mapsto e_i(\cdot;\gamma)$ is
a lifting of $\gamma$ generated by $e_i$.

	It is a simple exercise to verify that the components of $L$ and $D$
uniquely define (locally) their action on $u=u^ie_i(s;\gamma)$ and
$\lambda\in\PLift^1(E,\pi,M)$,
$\lambda\colon\gamma\mapsto\lambda_\gamma=\lambda_\gamma^i\hat{e}_i$,
according to
	\begin{gather}	\label{3.9}
L_{s\to t}^{\gamma} u
	=:\Sprindex[L]{j}{i}(t,s;\gamma) u^j e_i(t;\gamma)
\\			\label{3.10}
D_s^\gamma \lambda
	=: \Bigl(
	   \frac{\od\lambda_\gamma^i(s)}{\od s}
	   +
	   \Sprindex[\Gamma]{j}{i}(s;\gamma) \lambda_\gamma^j(s)
	   \Bigr) e_i(s;\gamma)
	\end{gather}
and that a change
$\{e_i(s;\gamma)\} \mapsto \{e'_i(s;\gamma)=A_i^j(s;\gamma)e_j(s;\gamma)\}$
with a non\ndash degenerate matrix\ndash valued function
 $A(s;\gamma) := [A_i^j(s;\gamma)]$ implies the transformation
	\begin{gather}	\label{3.11}
\Mat{L}(t,s;\gamma)\mapsto\Mat{L}^\prime(t,s;\gamma)
  = A^{-1}(t;\gamma) \Mat{L}(t,s;\gamma) A(s;\gamma)
\displaybreak[1]\\	\label{3.12}
\Mat{\Gamma}^\prime(s;\gamma)
  = A^{-1}(s;\gamma) \Mat{\Gamma}(s;\gamma) A(s;\gamma)
    + A^{-1}(s;\gamma)\frac{\od A(s;\gamma)}{\od s}.
	\end{gather}

	A crucial role further will be played by the \emph{coefficients}
$\Sprindex[\Gamma]{j}{i}(t,s;\gamma)$ in a frame $\{e_i\}$ of  linear
transport $L$,
	\begin{equation}	\label{3.13}
\Sprindex[\Gamma]{j}{i}(s;\gamma)
   := \frac{\pd\Sprindex[L]{j}{i}(s,t;\gamma)}{\pd t}\bigg|_{t=s}
   = - \frac{\pd\Sprindex[L]{j}{i}(s,t;\gamma)}{\pd s}\bigg|_{t=s} .
	\end{equation}
The usage of the same notation for the \emph{coefficients} of a
transport $L$ and \emph{components} of derivation $D$ along paths is not
accidental and finds its reason in the next fundamental
result~\cite[sec.~2]{bp-normalF-LTP}. Call a transport $L$ differentiable of
call $C^k$, $k=0,1,\dots$ if its matrix $\Mat{L}(t,s;\gamma)$ has $C^k$
dependence on $t$ (and hence on $s$ --- see~\cite[sec.~2]{bp-normalF-LTP}).
\emph{Every $C^1$ linear transport $L$ along paths generates a derivation $D$
along paths via
	\begin{equation}	\label{3.14}
D_{s}^{\gamma}(\lambda)
   := \lim_{\varepsilon\to0}\Bigl\{\frac{1}{\varepsilon}\bigl[
       L_{s+\varepsilon\to s}^{\gamma}\lambda_\gamma(s+\varepsilon)
       - \lambda_\gamma(s)\bigr]\Bigr\}
	\end{equation}
for every lifting $\lambda\in\PLift^1(E,\pi,B)$
with $\lambda\colon\gamma\mapsto\lambda_\gamma$
and conversely, for any derivation $D$ along paths there exists a unique
linear transport along paths generating it via~\eref{3.14}.}
Besides, if $L$ and $D$ are connected via~\eref{3.14}, the coefficients of
$L$ coincide with the components of $D$. In short, there is a bijective
correspondence between linear transports and derivations along paths given
locally through the equality of their coefficients and components
respectively.

	More details and results on the above items can be found
in~\cite{bp-normalF-LTP}.


\section
[Links between linear connections and linear transports]
{Links between linear connections and\\ linear transports}
\label{Sect4}

	Suppose $\gamma\colon J\to M$ is a  $C^1$ path and $\dot\gamma(s)$,
$s\in J$, is the vector tangent to $\gamma$ at $\gamma(s)$ (more precisely, at
$s)$. Let $\nabla$ and $D$ be, respectively, a linear connection and
derivation along paths in vector bundle $(E,\pi,M)$ and in a pair of frames
$(\{e_i\},\{E_\mu\})$ over some open set in $M$ the coefficients of $\nabla$
and the components of $D$ be $\Sprindex[\Gamma]{j\mu}{i}$ and
$\Sprindex[\Gamma]{j}{i}$ respectively, \ie
 $\nabla_{E_\mu} = \Sprindex[\Gamma]{i\mu}{j} e_j$
and
 $D_s^\gamma \hat{e}_i = \Sprindex[\Gamma]{i}{j} e_j(\gamma(s)$
with
$\hat{e}_i\colon\gamma\mapsto\hat{e}_i|_\gamma\colon s\mapsto e_i(\gamma(s))$
being lifting of paths generated by $e_i$. If
$\sigma=\sigma^ie_i\in\Sec^1(E,\pi,M)$ and $\hat\sigma\in\PLift(E,\pi,M)$ is
given via
$\hat\sigma\colon\gamma\mapsto\hat\sigma_\gamma:=\sigma\circ\gamma$,
then~\eref{3.10} implies
	\begin{equation*}
D^\gamma_s \hat\sigma
= \Big(
	\frac{\od\sigma^i(\gamma(s))}{\od s}
	+
	\Sprindex[\Gamma]{j}{i}(s;\gamma) \sigma^j(\gamma(s))
  \Bigr) e_i(\gamma(s))
	\end{equation*}
while, if $\gamma(s)$ is not a self-intersection point for
$\gamma$,~\eref{2.2} leads to
	\begin{equation*}
(\nabla_{\dot\gamma}\sigma)|_{\gamma(s)}
= \Big(
	\frac{\od\sigma^i(\gamma(s))}{\od s}
	+
	\Sprindex[\Gamma]{j\mu}{i}(\gamma(s))
  					\sigma^j(\gamma(s))\gamma^\mu(s)
  \Bigr) e_i(\gamma(s)) .
	\end{equation*}
Obviously, we have
	\begin{equation}	\label{4.1}
D^\gamma_s \hat\sigma = (\nabla_{\dot\gamma}\sigma)|_{\gamma(s)}
	\end{equation}
for every $\sigma$ iff
	\begin{equation}	\label{4.2}
\Sprindex[\Gamma]{j}{i}(s;\gamma)
=
\Sprindex[\Gamma]{j\mu}{i}(\gamma(s)) \dot\gamma^\mu(s)
	\end{equation}
which, in matrix form reads
	\begin{equation}	\label{4.3}
\Mat{\Gamma}(s;\gamma) = \Gamma_\mu(\gamma(s))\dot\gamma^\mu(s) .
	\end{equation}
A simple algebraic calculation shows that this equality is invariant under
changes of the frames $\{e_i\}$ in $(E,\pi,M)$ and $\{E_\mu\}$ in
$(T(M),\pi_T,M)$. Besides, if~\eref{4.2} holds, then $\Mat{\Gamma}$
transforms according to~\eref{3.12} iff $\Gamma_\mu$ transforms according
to~\eref{2.5}.

	The above considerations are a hint that the linear connections
should, and in fact can, be described in terms of derivations or,
equivalently, linear transports along paths; the second description being
more relevant if one is interested in the parallel transports generated by
connections.

	\begin{Thm}	\label{Thm4.1}
	If $\nabla$ is a linear connection, then there exists a derivation $D$
along paths such that~\eref{4.1} holds for every $C^1$ path
$\gamma\colon J\to M$ and every $s\in J$ for which $\gamma(s)$ is not
self\ndash intersection point for $\gamma$.%
\footnote{%
In particular, $\gamma$ can be injective and $s$ arbitrary. If we restrict
the considerations to injective paths, the derivation $D$ is unique. The
essential point here is that at the self\ndash intersection points of
$\gamma$, if any, the mapping
$\dot\gamma\colon\gamma(s)\mapsto\dot\gamma(s)$
is generally multiple\ndash valued and, consequently, it is not a vector
field (along $\gamma$); as a result
$(\nabla_{\dot\gamma}\sigma)|_{\gamma(s)}$ at them becomes also
multiple\ndash valued.%
}
The matrix of the components of $D$ is given by~\eref{4.3} for every $C^1$
path $\gamma\colon J\to M$ and  $s\in J$ such that $\gamma(s)$ is not a
self\ndash intersection point for $\gamma$. Conversely, given a derivation
$D$ along path whose matrix along any $C^1$ path $\gamma\colon J\to M$ has
the form~\eref{4.3} for some matrix\ndash valued functions $\Gamma_\mu$,
there is a unique linear connection $\nabla$ whose matrices of coefficients
are exactly $\Gamma_\mu$ and for which, consequently,~\eref{4.1} is valid at
the not self\ndash intersection points of $\gamma$.
	\end{Thm}

	\begin{Proof}
	NECESSITY.
	If $\Gamma_\mu$ are the matrices of the coefficients of $\nabla$ in
some pair of frames $(\{e_i\},\{E_\mu\})$, define the matrix $\Mat{\Gamma}$
of the components of $D$ via~\eref{4.3} for any $\gamma\colon J\to M$.
	SUFFICIENCY. Given $D$ for which the decomposition~\eref{4.3} holds
in $(\{e_i\},\{E_\mu\})$ for any $\gamma$. It is trivial to verity that
$\Gamma_\mu$ transform according to~\eref{2.5} and, consequently, they are
the matrices of the coefficients of a linear connection $\nabla$ for which,
evidently,~\eref{4.1} holds.
	\end{Proof}

	A trivial consequence of the above theorem is the next important
result.

	\begin{Cor}	\label{Cor4.1}
	There is a bijective correspondence between the set of linear
connections in a vector bundle and the one of derivations along paths in it
whose components' matrices admit (locally) the decomposition~\eref{4.3}.
Locally, along a $C^1$ path $\gamma$ and pair of frames $(\{e_i\},\{E_\mu\})$
along it, it is given by~\eref{4.3} in which $\Mat{\Gamma}$ and $\Gamma_\mu$
are the matrices of the components of a derivation along paths and of the
coefficients of a linear connection respectively.
	\end{Cor}

	Let us now look on the preceding material from the view-point of
linear transports along paths and parallel transports generated by linear
connections.

	Recall (see, e.g.,~\cite[chapter~2]{Poor}), a section
$\sigma\in\Sec^1(E,\pi,M)$ is parallel along  $C^1$ path $\gamma\colon J\to
M$ with respect to a linear connection $\nabla$ if
$(\nabla_{\dot\gamma}\sigma)|_{\gamma(s)}=0$, $s\in J$.%
\footnote{%
If $\gamma$ is not injective, here and henceforth
$(\nabla_{\dot\gamma}\sigma)|_{\gamma(s)}$
should be replaced by $D_s^\gamma\hat\sigma$,
$\hat\sigma\colon\gamma\mapsto\sigma\circ\gamma$, where $D$ is the derivation
along paths corresponding to $\nabla$ via corollary~\eref{Cor4.1}.%
}
The parallel transport along a $C^1$ path $\alpha\colon[a,b]\to M$,
$a,b\in\field[R]$, $a\le b$, generated by $\nabla$ is a mapping
\[
P^\alpha\colon \pi^{-1}(\alpha(a))\to \pi^{-1}(\alpha(b))
\]
such that $P^\alpha(u_0):=u(b)$ for every element $u_0\in\pi^{-1}(\alpha(a))$
where $u\in\Sec^1(E,\pi,M)|_{\alpha([a,b])}$ is the unique solution of the
initial\ndash value problem
	\begin{equation}	\label{4.4}
\nabla_{\dot\alpha} u = 0,
\qquad
u(a)=u_0 .
	\end{equation}
The parallel transport $P$ generated by (assigned to, corresponding to) a
linear connection $\nabla$ is a mapping assigning to any
$\alpha\colon[a,b]\to M$ the parallel transport $P^\alpha$ along $\alpha$
generated by $\nabla$.

	Let $D$ be the derivation along paths corresponding to $\nabla$
according to corollary~\eref{Cor4.1}. Then~\eref{4.1} holds for
$\gamma=\alpha$, so~\eref{4.4} is tantamount to
	\begin{equation}	\label{4.5}
D_s^\alpha \hat u = 0
\qquad
u(a) = u_0
	\end{equation}
where $\hat u\colon\alpha\mapsto \bar u\circ\alpha$ with $\bar
u\in\Sec^1(E,\pi,M)$ such that $\bar u|_{\alpha([a,b])}=u$. From here and the
results of~\cite[sec.~2]{bp-normalF-LTP} immediately follows that the lifting
$\hat u$ is generated by the unique linear transport $\Psf$ along paths
corresponding to  $D$,
	\begin{equation}	\label{4.6}
\hat u\colon\alpha\mapsto
\hat{u}_\alpha := \bar{\Psf}_{a,u_0}^{\alpha},
\quad
\bar{\Psf}_{a,u_0}^{\alpha}\colon s\mapsto
\bar{\Psf}_{a,u_0}^{\alpha}(s):= \Psf_{a\to s}^{\alpha} u_0,
\qquad
s\in [a,b] .
	\end{equation}
Therefore
\(
P^\alpha(u_o)
:= u(b)
= \bar u(\alpha(b))
= \hat{u}_\alpha(b)
= \Psf_{a\to b}^{\alpha} u_0 .
\)
Since this is valid for all $u_0\in\pi^{-1}(\alpha(a))$, we have
	\begin{equation}	\label{4.7}
P^\alpha = \Psf_{a\to b}^{\alpha} .
	\end{equation}

	\begin{Thm}	\label{Thm4.2}
	The parallel transport $P$ generated by a linear connection $\nabla$
in a vector bundle coincides, in a sense of~\eref{4.7}, with the unique linear
transport $\Psf$ along paths in this bundle corresponding to the derivation
$D$ along paths defined, via corollary~\ref{Cor4.1}, by the connection.
Conversely, if $\Psf$ is a linear transport along paths whose coefficients'
matrix admits the representation~\eref{4.1}, then for every $s.t\in[a,b]$
	\begin{equation}	\label{4.8}
\Psf_{s\to t}^{\alpha} =
\begin{cases}
P^{\alpha|[s,t]}			&\text{for $s\le t$}	\\
\bigl(P^{\alpha|[t,s]}\bigr)^{-1}	&\text{for $s\ge t$}
\end{cases} ,
	\end{equation}
where $P$ is the parallel transport along paths generated by the unique
linear connection $\nabla$ corresponding to the derivation $D$ along paths
defined by $\Psf$.
	\end{Thm}

	\begin{Proof}
	The first part of the assertion was proved above while
deriving~\eref{4.7}. The second part is simply the inversion of all logical
links in the first one, in particular~\eref{4.8} is the solution
of~\eref{4.7} with respect to $\Psf$.
	\end{Proof}

	\begin{Rem}	\label{Rem4.1}
	In all of the above results a crucial role plays the (local)
condition~\eref{4.3}. It has also an invariant version in terms of linear
transports: for a given transport $L$ it is ``almost'' equivalent to the
conditions
\(
L_{s\to t}^{\gamma\circ\varphi}
=
L_{\varphi(s)\to \varphi(t)}^{\gamma},
\)
 $s,t\in J^{\prime\prime}$,
and
\(
L_{s\to t}^{\gamma|J'} = L_{s\to t}^{\gamma},
\)
 $s,t\in J'$,
where $\gamma\colon J\to M$, $\varphi\colon J^{\prime\prime}\to J$ is
orientation\ndash preserving diffeomorphism, and $J'\subseteq J$ is a
subinterval. This means that, in some sense, a linear transport $L$ is a
parallel one (generated by a linear connection) iff it satisfies these
conditions. For details, see~\cite{bp-TP-parallelT}. A revised and expanded
study of the links between linear and parallel transports will be given
elsewhere.
	\end{Rem}

	The transport $\Psf$ along paths corresponding according to
theorem~\ref{Thm4.2} to a parallel transport $P$ or a linear connection
$\nabla$ will be called \emph{parallel transport along paths}.

	\begin{Cor}	\label{Cor4.2}
	The local coefficients' matrix $\Gamma$ of a parallel transport along
paths and the coefficients' matrices $\Gamma_\mu$ of the generating it (or
generated by it) linear connection are connected via~\eref{4.3} for every
$C^1$ path $\gamma\colon J\to M$.
	\end{Cor}

	\begin{Proof}
	See theorem~\ref{Thm4.2}.
	\end{Proof}


\section {Frames normal for linear connections}
\label{Sect5}

	In the series of
papers~\cite{bp-Frames-n+point,bp-Frames-path,bp-Frames-general} the problems
of existence, uniqueness, and holonomicity of frames normal for derivations
of the tensor algebra over a manifold were completely solved on arbitrary
submanifolds. In particular, all of these results apply for linear
connections on manifolds, \ie for linear connections in the tangent bundle
over a manifold. The purpose of this section is to be obtained similar
results for linear connections in arbitrary finite\ndash dimensional vector
bundles whose base and bundle spaces are $C^\infty$ manifolds. The method we
are going to follow is quite simple: relying on the conclusions of the
previous sections, we shall transfer the general results
of~\cite{bp-normalF-LTP} concerning frames normal for linear transports to
analogous ones regarding linear connections. More precisely, the methods of
sections~5--7 of~\cite{bp-normalF-LTP} should be applied as~\eref{4.3} holds
for parallel transports generated by linear connections. Equivalently well,
as we shall see, the methods and results
of~\cite{bp-Frames-n+point,bp-Frames-path,bp-Frames-general} can almost
directly be used.

	\begin{Defn}	\label{Defn5.1}
	Given a linear connection $\nabla$ in a vector bundle $(E,\pi,M)$ and
a subset $U\subseteq M$. A frame $\{e_i\}$ in $E$ defined over an open subset
$V$ of $M$ containing $U$ or equal to it, $V\supseteq U$, is called normal
for $\nabla$ over $U$ if in it and some (and hence any) frame $\{E_\mu\}$ in
$T(M)$ over $V$ the coefficients of $\nabla$ vanish everywhere on $U$.
Respectively, $\{e_i\}$ is normal for $\nabla$ along a mapping $g\colon Q\to
M$, $Q\not=\varnothing$, if $\{e_i\}$ is normal for $\nabla$ over $g(Q)$.
	\end{Defn}

	If one wants to attack directly the problems for existence,
uniqueness, \etc of frames normal for a linear connection $\nabla$, the
transformation formula~\eref{2.5} should be used. Indeed, if
$(\{e_i\},\{E_\mu\})$ is an arbitrary pair of frames over $V\supseteq U$, a
frames $\{e'_i=A_i^je_j\}$ is normal for $\nabla$ over $U$ if for some
$\{E'_\mu=B_\mu^\nu E_\nu\}$ in the pair $(\{e'_i\},\{E'_\mu\})$ is
fulfilled $\Sprindex[\Gamma]{j\mu}{\prime\,i}|_U=0$, which, by~\eref{2.5}, is
equivalent to
	\begin{equation}	\label{5.1}
\bigl( \Gamma_\nu A + E_\nu(A) \bigr) \big|_U = 0 .
	\end{equation}
We call this (matrix) equation the \emph{equation of the normal frames} for
$\nabla$ over $U$ or simply the \emph{normal frames equation} (for $\nabla$
on $U$). It contains all the information for the frames normal for a given
linear connection, if any. Since in~\eref{5.1} the matrix $B=[B_\mu^\nu]$
does not enter, the trivial but important corollary of it is that the choice
of the frame $\{E_\mu\}$ over $V$ in $T(M)$ is completely insignificant in a
sense that if in $(\{e'_i\},\{E'_\mu\})$ the coefficients of $\nabla$ vanish
on $U$, then they also have this property in
$(\{e'_i\},\{E^{\prime\prime}_\mu\})$ for any other frame
$\{E^{\prime\prime}_\mu\}$ over $V$ in $T(M)$.

	If one likes, he/she could  begin an independent investigation of the
normal frames equation~\eref{5.1} with respect to the $C^1$ non\ndash
degenerate matrix\ndash valued function $A$ which performs the transition
from an arbitrary fixed (chosen) frame $\{e_i\}$ to normal ones, if any.
But we are not going to do so since this equation has been completely studied
in the practically most important (at the moment) cases, the only thing
needed is the existing results to be carried across to linear connections.

	Recall~\cite[definition~7.2]{bp-normalF-LTP}, a frame $\{e_i\}$ is
called strong normal on $U$ for a linear transport $L$ along paths, for
which~\eref{4.3} holds, if in $(\{e_i\},\{E_\mu\})$ for some frame
$\{E_\mu\}$ the \emph{3\ndash index coefficients' matrices}  $\Gamma_\mu$ of
$L$ vanish on $U$.
	\begin{Prop}	\label{Prop5.1}
	The frames normal for a linear connection in a vector bundle are
strong normal for the corresponding to it parallel transport along paths and
vice versa.
	\end{Prop}
	\begin{Proof}
	See corollary~\ref{Cor4.2}.
	\end{Proof}

	As we pointed in~\cite[sec.~7]{bp-normalF-LTP}, the most interesting
problems concerning strong normal frames are practically solved
in~\cite{bp-Frames-n+point,bp-Frames-path,bp-Frames-general}. Let us repeat
the arguments for such a conclusion and state, due to
proposition~\ref{Prop5.1}, the main results in terms of linear connections.

	Let $(E,\pi,M)$ be finite-dimensional vector bundle with $E$ and $M$
being $C^\infty$ manifolds, $U\subseteq M$, $V\subseteq M$ be an \emph{open}
subset containing $U$, $V\supseteq U$, and $\nabla$ be linear connection in
$(E,\pi,M)$. The problem is to be investigated the frames normal for $\nabla$
over $U$ or, equivalently, the ones strong normal for the parallel transport
along paths generated by $\nabla$.

	Above we proved that a frame $\{e'_i\}$ over $V$ in $E$ is normal for
$\nabla$ over $U$ if and only if for arbitrarily fixed pair of frames
$(\{e_i\},\{E_\mu\})$, $\{e_i\}$ in $E$ and $\{E_\mu\}$ in $T(M)$, over $V$
there is a non\ndash degenerate  $C^1$ matrix\ndash valued function $A$
satisfying~\eref{5.1}, in which $\Gamma_\mu$ are the coefficients' matrices
of $\nabla$ in $(\{e_i\},\{E_\mu\})$, and such that $e'_i=A_i^je_j$. In
other words, $\{e'_i\}$ is normal for $\nabla$ over $U$ if it can be
obtained from an arbitrary frame $\{e_i\}$ via transformation whose matrix
is a solution of~\eref{5.1}.

	Comparing equation~\eref{5.1} with analogous ones
in~\cite{bp-Frames-n+point,bp-Frames-path,bp-Frames-general}, we see that
they are identical with the only difference
that the size of the square matrices $\Gamma_1,\ldots,\Gamma_{\dim M}$, and
$A$
in~\cite{bp-Frames-n+point,bp-Frames-path,bp-Frames-general}
is $\dim M\times\dim M$ while in~\eref{5.1} it is
$v\times v$, where $v$  is the dimension of the vector bundle $(E,\pi,M)$,
\ie $v=\dim\pi^{-1}(x)$, $x\in M$, which is generally not equal to $\dim M$.
But this difference is completely insignificant from the view\ndash point of
solving these equations (in a matrix form) or with respect to the
integrability conditions for them. Therefore all of the results
of~\cite{bp-Frames-n+point,bp-Frames-path,bp-Frames-general},
concerning the solution of the matrix
differential equation~\eref{5.1}, are (\emph{mutatis mutandis}) applicable
to the investigation of the frames strong normal on a set $U\subseteq M$.

	The transferring of the results
from~\cite{bp-Frames-n+point,bp-Frames-path,bp-Frames-general}
is so trivial that their explicit reformulations has a sense if one really
needs the corresponding rigorous assertions for some concrete purpose. By
this reason, we want to describe below briefly the general situation and one
its corollary.

	\begin{Thm}[see \protect{\cite[theorem~3.1]{bp-Frames-general}}]
	\label{Thm5.1}
	If $\gamma_n\colon J^n\to M$, $J^n$ being neighborhood in
$\mathbb{R}^n$, $n\in\mathbb{N}$, is a $C^1$ injective mapping,
then a necessary and sufficient
condition for the existence of frame(s) normal over $\gamma_n(J^n)$
for some linear connection is, in some neighborhood (in $\mathbb{R}^n$) of
every $s\in J^n$, their (3\ndash index) coefficients to satisfy the equations
	\begin{equation}	\label{5.2}
\bigl(
  R_{\mu\nu} (-\Gamma_1\circ\gamma_n,\ldots,-\Gamma_{\dim M}\circ\gamma_n)
\bigr)
	(s)
  = 0,
\qquad \mu,\nu=1,\ldots,n
	\end{equation}
where $R_{\mu\nu}$
(in a coordinate frame
$\bigl\{E_\mu=\frac{\pd}{\pd x^\mu}\bigr\}$ in a neighborhood of $x\in M$)
are given via
	\begin{equation*}
R_{\mu\nu}(-    {\Gamma}_1,\ldots,-    {\Gamma}_{\dim M})
  := - \frac{\pd    {\Gamma}_\mu}{\pd x^\nu}
     + \frac{\pd    {\Gamma}_\nu}{\pd x^\mu}
     +     {\Gamma}_\mu    {\Gamma}_\nu
     -     {\Gamma}_\nu    {\Gamma}_\mu .
	\end{equation*}
for $x^\mu=s^\mu$, $\mu,\nu=1,\dots,n$ with $\{s^\mu\}$ being Cartesian
coordinates in $\mathbb{R}^n$.
	\end{Thm}

	From~\eref{5.2} an immediate observation
follows~\cite[sect.~6]{bp-Frames-general}: strong normal frames always exist
at every point ($n=0$) or/and along every $C^1$ injective path ($n=1$).
Besides, these are
the \emph{only cases} when normal frames \emph{always exist} because for them
~\eref{5.2} is identically valid. On submanifolds with dimension greater
than or equal to two normal frames exist only as an exception if (and only
if)~\eref{5.2} holds. For $n=\dim M$ equations~\eref{5.2} express the
flatness of the corresponding linear connection.

	If on $U$ exists a frame $\{e_i\}$ normal for $\nabla$, then all
frames  $\{e'_i=A_i^je_j\}$ which are normal over $U$ can easily be
described: for the normal frames, the matrix $A=[A_i^j]$ must be such that
$E_\mu(A)|_U=0$ for some (every) frame $\{E_\mu\}$ over $U$ in $T(M)$
(see~\eref{5.1} with $\Gamma_\mu|_U=0$).

	These conclusions completely agree with the ones made
in~\cite[sec.~8]{bp-normalF-LTP} concerning linear connections on a manifold
$M$, \ie in the tangent bundle $(T(M),\pi_T,M)$.


\section
[Inertial frames and equivalence principle in gauge theories]
{Inertial frames and \\ equivalence principle in gauge theories}
\label{Sect6}

	In~\cite{bp-PE-P?} it was demonstrated that, when gravitational fields
are concerned, the inertial frames for them are the normal ones for the linear
connection describing the field and they coincide with the (inertial) frames
in which special theory of relativity is valid. The last assertion is the
contents of the (strong) equivalence principle. In the present section, relying
on the ideas at the end of~\cite[sec.~5]{bp-PE-P?}, we intend to transfer
these conclusions to the area of classical gauge theories.%
\footnote{%
The primary role of the principle of equivalence is to ensure the transition
from general to special relativity.  It has quite a number of versions, known
as weak and strong equivalence
principles~\cite[pp.~72--75]{Ivanenko&Sardanashvily-1985}, any one of which has
different, sometimes non\ndash equivalent, formulations. In the present paper
only the strong(est) equivalence principle is considered. Some of its
formulations can be found in~\cite{bp-PE-P?}.%
}

	Freely speaking, an inertial frame for a physical system is a one in
which the system behaves in some aspects like a free one, \ie such a frame
`imitates' the absence (vanishment) of some forces acting on the system.
Generally inertial frames exist only locally, \eg along injective paths, and
their existence does not mean the vanishment of the field responsible for a
particular force. The best known example of this kind of frames, as we
pointed above, is the gravitational field. Below we rigorously generalize
these ideas to all gauge fields.

	The gauge fields were introduced in connection with the study of
fundamental interactions between elementary particles.%
\footnote{%
See, e.g., the collection of papers~\cite{CompensationFields}.%
}
Later it was realized~\cite{Konopleva&Popov,Bleecker,Baez&Muniain} that, from
mathematical view\ndash point, they are equivalent to the concept of (linear)
connection on (principal) vector bundle which was clearly formulated a bit
earlier. The present day understanding is that%
\footnote{%
The next citation is from~\cite[p.~118]{Drechsler&Mayer}.%
}
``a gauge field is a connection on the principal fibration in which the
vector bundle of the particle fields is associated. More precisely, we
identify a gauge field with the connection 1\ndash form or with its
coefficients in terms of a local basis of the cotangent bundle of the base
manifold.''
Before proceeding on with our main topic, we briefly comment on this
definition of a gauge field.

	The definition of a principal bundle (fibre bundle, fibration) and
the associated with it vector bundle can be found in any serious book on
differential geometry or its applications
--- \eg in~\cite[ch.~I, \S~5]{K&N-1}, \cite[pp.~193--204]{Greub&et_al.-2}
or~\cite[p.26]{Bleecker} ---
and will not be reproduced here. A main feature of a principal bundle
$(P,\pi,M,G)$, consisting of a bundle $(P,\pi,M)$ and a Lie group $G$, is
that the (typical) fibre of $(P,\pi,M)$ is $G$ and $G$ acts freely on $P$ to
the right.

	Recall~\cite{K&N-1,Bleecker}, a connection 1\ndash form (of a linear
connection) is a 1\ndash form with values in the Lie algebra of the group
$G$, but, for the particular case and purposes, it can be considered a
matrix\ndash valued 1\ndash form, as it is done
in~\cite[p.~118]{Drechsler&Mayer} (cf.~\cite[ch.~III, \S~7]{K&N-1}). Let
$(E,\pi_E,M,F)$ be the vector bundle with fibre $F$ associated with
$(P,\pi,M,G)$ and some (left) action $L$ of $G$ on the manifold $F$.%
\footnote{%
See, e.g.,~\cite{Poor} or~\cite{Greub&et_al.-2} for details. In the physical
applications $F$ is a vector space and $L$ is treated as a representation of
$G$ on $F$, \ie a homomorphism $L\colon G\to\GL(F)$ from $G$ in the group
$\GL(F)$ of non\ndash degenerate linear mappings $F\to F$.%
}
According to known definitions and results, which, for instance, can be found
in~\cite{Poor,Greub&et_al.,Bleecker}, a connection 1\ndash form on
$(P,\pi,M,G)$  induces a linear connection (more precisely, covariant
derivative operator) $\nabla$ in the associated vector bundle $(E,\pi_E,M,F)$
in which the particle fields `live' as sections.%
\footnote{~%
The explicit construction of $\nabla$ can be found
in~\cite[p.245ff]{Pham-Mau-Quan}.%
}
The local coefficients
$\Sprindex[A]{j\mu}{i}$ of $\nabla$ in some pair of frames
$(\{e_i\},\{E_\mu\})$, $\{e_i\}$ in $E$ and $\{E_\mu\}$ in $T(M)$, represent
(locally) the connection 1\ndash form (gauge field) and are known as
\emph{vector potentials} in the physical literature.%
\footnote{%
For
example, see~\cite{Konopleva&Popov, Baez&Muniain, Bleecker}.
Often~\cite{Drechsler&Mayer,Slavnov&Fadeev} a particle field $\psi$ is
represented as a vector-colon (in a given frame $\{e_i\}$) transforming under
a representation $L(G)$ of the structure group $G$ in the group
$\GL(n,\field)$ of non\ndash degenerate $n\times n$, $n=\dim F$, matrices over
$\field=\field[R],\field[C]$. In this case the matrices
$A_\mu=[\Sprindex[A]{j\mu}{i}]_{i,j=1}^{n}$  are written as
$A_\mu=A_\mu^iT_i$ where $T_i$ are the matrices (generators) forming a basis
of the Lie algebra in the representation $L(G)$ and the function $A_\mu^i$
are known as Yang-Mills fields (if the connection satisfies the Yang-Mills
equation) which are also identified with the initial gauge field. Sometimes
the matrices $A_\mu$ are called Yang-Mills fields too and are considered as
(components of) a vector field with values in the Lie algebra of $L(G)$.%
}
Consequently, locally a gauge field can be identified with the vector
potentials which are the coefficients of the linear connection $\nabla$ (in
the associated bundle $(E,\pi_E,M,F)$) representing the gauge field.

	Relying on the previous experience with gravity~\cite{bp-PE-P?}, we
\emph{define the physical concept inertial frame for a gauge field to coincide
with the mathematical one normal frame for the linear connection whose local
coefficients (vector potentials) represent (locally) the gauge field}.
 This completely agrees with the said at the beginning of the present
section: according to the accepted procedure~\cite{Slavnov&Fadeev,Bleecker},
the Lagrangian of a particle field interacting with a gauge field is obtained
from the one of the same field considered as a free one by replacing the
ordinary (partial) derivatives with the covariant ones corresponding to the
connection $\nabla$ representing the gauge field. Therefore, in a frame
inertial on a subset $U\subseteq M$ for a gauge field the Lagrangian of a
particle field interacting with the gauge field coincides with the Lagrangian
for the same field considered as a free one.%
\footnote{%
Generally this does not mean that in an inertial frame disappear (all of) the
physical effects of the gauge field as they, usually, depend on the curvature
of describing it linear connection. Besides, it is implicitly supposed the
Lagrangians to depend on the particle fields via them and their first
derivatives.%
}
So, we can assert that in an inertial frame the physical effects depending
directly on gauge field (but not on its derivatives!) disappear. From the
results obtained in the present work directly follows the existence of
inertial frames for a gauge field at any fixed spacetime point or/and along
injective path. On other subsets of the spacetime inertial frames may exist
only as an exception for some particular gauge fields.%
\footnote{%
On submanifolds these special fields are selected by theorem~\ref{Thm5.1}.%
}

	The analogy with gravity is quite clear and it is due to the simple
fact that the gravitational as well as gauge fields are locally described via
the local coefficients of linear connections, in the bundle tangent to the
spacetime in the former case and in some other bundle over it in the latter
one. This state of affairs can be pushed further. The above\ndash mentioned
procedure for getting the non\ndash free Lagrangian (or field equations) for
a particle field interacting with a gauge one is nothing else than the
\emph{minimal coupling (replacement, interaction) principle} applied to the
particular situation.
As a result the free Lagrangian (or field equation) plays the role
of a Lagrangian (field equation) in an inertial frame in the sense of special
relativity~\cite{bp-PE-P?}. Call a frame $\{e_i\}$, in the bundle space of
the bundle associated with the principal bundle in which particle fields
live, \emph{inertial} (in a sense of special relativity) if in it the field
Lagrangian (equation) is free one. Now we can formulate the \emph{equivalence
principle} in gauge field theories (cf.~\cite[p.~216]{bp-PE-P?}). It assets
the coincidence of two types of inertial frames:
\emph{the normal ones in which the vector potentials of a gauge field
(considered as linear connection) vanish and the inertial frames in which the
Lagrangian (field equation) of a particle field interacting with the gauge
one is free}.%
\footnote{%
Notice, here and above we do not suppose the spacetime to be flat.%
}
According to the above discussion
\emph{the equivalence principle is a theorem},
not an axiom, in gauge theories as one can expected from a similar result in
gravity.

	Consequently, we have a separate equivalence principle for each gauge
field. Can we speak of a single equivalence principle concerning
simultaneously \emph{all gauge fields and gravity}? The answer is expected
(in a sense) to be positive. However its argumentation and explanation
depends on the particular theory one investigates since at this point we meet
the problem of unifying the fundamental interactions describe mathematically
via linear connections in vector bundles. Below we outline the most simple
situation, which can be called a `direct sum of the interactions' and does
not predict new physical phenomena but on its base one may do further
research on the subject.

	Suppose a particle field  $\psi$ interacts with (independent) gauge
fields represented as linear connections
$\nabla^{(a)}$, $a=1,\dots,n$, $n\in\field[N]$,
acting in vector bundles $\xi_a:=(E_a,\pi_a,M,)$, $a=1,\dots,n$,
respectively with $M$ being a manifold used as spacetime model. To include
the gravity in the scheme, we assume it to be describe by a (possibly
(pseudo\ndash)Riemannian) linear connection $\nabla^{(0)}$ on $M$, \ie in the
tangent bundle $\xi_0:=(T(M),\pi_T,M)=(E_0,\pi_0,M)$.
Let $\xi:=(E,\pi,M\times\dotsb\times M)$, where $M$ is taken $n+1$ times, be
the direct sum~\cite{Poor,Baez&Muniain} of the bundles $\xi_0,\dots,\xi_n$.%
\footnote{%
For purposes which will be explained elsewhere the direct sum of the mentioned
bundles should be replace with the bundle
$(\Esf,\bs{\pi},M)$
where
\(
\Esf:=\{
	(u_0,\dots,u_n)\in E_0\times\dotsb\times E_n
	:
	\pi_0(u_0)=\dots=\pi(u_n)
      \}
\)
and $\bs{\pi}(u_0,\dots,u_n) := \pi_0(u_0)$
for $(u_0,\dots,u_n)\in E$. So that
$ \pi^{-1}(x)=\pi^{-1}_0(x)\times\dots\times\pi^{-1}_n(x) $
for $x\in M$.%
}
In this case the particle field $\psi$ should be considered as a section
$\psi\in\Sec^2(\xi)$ and the system of gauge fields with which it interacts
is represented by a connection $\nabla$ equal to the direct sum of
 $\nabla^{(0)},\dots,\nabla^{(n)}$,
$\nabla=\nabla^{(0)}\times\dots\times\nabla^{(n)}$,
(see, e.g.,~\cite[p.~254]{Baez&Muniain}).

	Now the minimal coupling principle says that the non-free Lagrangian
of $\psi$ is obtained from the free one by replacing in it the partial
derivatives with covariant ones with respect to $\nabla$. Since the fields
with which $\psi$ interacts are supposed independent, the frames inertial for
them, if any, are completely independent. Therefore if for some set
$U\subseteq M$ and any $a=0,\dots,n$ there are frame
 $\{ e_{i_a}^{(a)} : i_a=1,\dots,\dim\pi^{-1}_a(x),\ x\in M \}$
normal over $U$ for $\nabla^{(a)}$, the direct product of these frames,
\(
\{e_i : =1,\dots\dim\pi^{-1}(x),\ x\in M \}
:=
\{ e_{i_0}^{(0)}\times\dots\times e_{i_n}^{(n)} \},
\)
is a frame normal over $U$ for $\nabla$. In this sense $\{e_i\}$ is an
inertial frame for the considered system of fields. We can assert the
existence of such frames at any point in $M$ and/or along any injective path
in $M$. Now the principle of equivalence becomes the trivial assertion that
inertial frames for the system of fields coincide with the normal ones for
$\nabla$.


\section {Conclusion}
\label{Conclusion}

	The main result of this paper is that a number of important results
concerning existence, uniqueness, holonomicity, construction, \etc of frames
normal for linear transports (or derivations) along paths or derivations of
the tensor algebra over a manifold remain (\emph{mutatis mutandis}) valid for
linear connections in vector bundles. A particular example for that being
theorem~\ref{Thm5.1} from which follows that any linear connection in a
vector bundle admits frames normal at a single point or/and along an
injective path. As a consequence, as we saw, the concept of an inertial frame
(of reference), usually associated to systems in gravitational field, can be
transferred to the area of gauge theories which, in turn, allows the extension
of the range of validity of the principle of equivalence for gravitational
physics to systems interacting via gauge fields (and, of course,
gravitationally).

	We would like to say that the physical importance of the normal
frames, more precisely of normal coordinates, was notice in different
directions already in the early works on normal coordinates,
like~\cite{Pinl,ORai}.

	At the end, we shall mention the geometric equivalence principle
(see:~\cite[p.~76]{Ivanenko&Sardanashvily-1985},
      \cite[p.~19]{Ivanenko&Sardanashvily-1983},
      \cite[p.~3]{Sardanashvily&Zakharov-1991}):
there are reference frames with respect to which Lorentz invariants can be
defined everywhere on the spacetime and that are constant under parallel
transport. A possible item for further research is to replace here the Lorentz
invariants with the ones (of a representation) of the structure group of a
gauge theory which will lead to the transferring of the (geometric)
equivalence principle to the gauge theory whose structure group is involved.


\addcontentsline{toc}{section}{References}
\bibliography{bozhopub,bozhoref}

\begin{thebibliography}{10}

\bibitem{bp-Frames-n+point}
Bozhidar~Z. Iliev.
\newblock Normal frames and the validity of the equivalence principle: {I}.
  {Cases} in a neighborhood and at a point.
\newblock {\em Journal of Physics A: Mathematical and General},
  29(21):6895--6901, 1996.
\newblock LANL arXiv server, E-print No.\ gr-qc/9608019.

\bibitem{bp-Frames-path}
Bozhidar~Z. Iliev.
\newblock Normal frames and the validity of the equivalence principle: {II}.
  {The} case along paths.
\newblock {\em Journal of Physics A: Mathematical and General},
  30(12):4327--4336, 1997.
\newblock LANL arXiv server, E-print No.\ gr-qc/9709053.

\bibitem{bp-Frames-general}
Bozhidar~Z. Iliev.
\newblock Normal frames and the validity of the equivalence principle: {III}.
  {The} case along smooth maps with separable points of self-intersection.
\newblock {\em Journal of Physics A: Mathematical and General},
  31(4):1287--1296, January 1998.
\newblock LANL arXiv server, E-print No.\ gr-qc/9805088.

\bibitem{bp-Bases-n+point}
Bozhidar~Z. Iliev.
\newblock Special bases for derivations of tensor algebras. {I}.~{Cases} in a
  neighborhood and at a point.
\newblock JINR Communication E5-92-507, Dubna, 1992.
\newblock (see~\cite{bp-Frames-n+point}).

\bibitem{bp-Bases-path}
Bozhidar~Z. Iliev.
\newblock Special bases for derivations of tensor algebras. {II}.~{Case} along
  paths.
\newblock JINR Communication E5-92-508, Dubna, 1992.
\newblock (see~\cite{bp-Frames-path}).

\bibitem{bp-Bases-general}
Bozhidar~Z. Iliev.
\newblock Special bases for derivations of tensor algebras. {III}.~{Case} along
  smooth maps with separable points of selfintersection.
\newblock JINR Communication E5-92-543, Dubna, 1992.
\newblock (see~\cite{bp-Frames-general}).

\bibitem{bp-normalF-LTP}
Bozhidar~Z. Iliev.
\newblock Normal frames and linear transports along paths in vector bundles.
\newblock LANL arXiv server, E-print No.\ gr-qc/9809084, 1998.

\bibitem{K&N}
S.~Kobayashi and K.~Nomizu.
\newblock {\em Foundations of Differential Geometry}, volume I and II.
\newblock Interscience Publishers, New York-London-Sydney, 1963 and 1969.

\bibitem{Poor}
Walter~A. Poor.
\newblock {\em Differential geometric structures}.
\newblock McGraw-Hill Book Company Inc., New York, 1981.

\bibitem{Greub&et_al.}
W.~Greub, S.~Halperin, and R.~Vanstone.
\newblock {\em Connections, Curvature, and Cohomology}, volume 1,2,3.
\newblock Academic Press, New York and London, 1972, 1973, 1976.

\bibitem{Bleecker}
David Bleecker.
\newblock {\em Gauge theory and variational principles}.
\newblock Addison-Wiley Pub.~Co., London, 1981.

\bibitem{Baez&Muniain}
John Baez and Javier~P.\ Muniain.
\newblock {\em Gauge fields, knots and gravity}, volume~4 of {\em Series in
  knots and Everything}.
\newblock World Scientific, Singapore-New Jersey-London-Hong Kong, 1994.

\bibitem{K&N-1}
S.~Kobayashi and K.~Nomizu.
\newblock {\em Foundations of Differential Geometry}, volume~I.
\newblock Interscience Publishers, New York-London, 1963.

\bibitem{Drechsler&Mayer}
W.~Drechsler and M.~E. Mayer.
\newblock {\em Fibre bundle techniques in gauge theories}, volume~67 of {\em
  Lecture notes in physics}.
\newblock Springer-Verlag, Berlin-Heidelberg-New York, 1977.

\bibitem{bp-TP-parallelT}
Bozhidar~Z. Iliev.
\newblock Transports along paths in fibre bundles. {II}.~{Ties} with the theory
  of connections and parallel transports.
\newblock JINR Communication E5-94-16, Dubna, 1994.

\bibitem{bp-PE-P?}
Bozhidar~Z. Iliev.
\newblock Is the principle of equivalence a principle?
\newblock {\em Journal of Geometry and Physics}, 24(3):209--222, 1998.
\newblock LANL arXiv server, E-print No.\ gr-qc/9806062.

\bibitem{Ivanenko&Sardanashvily-1985}
Dmitri~D. Ivanenko and Gennadi~A. Sardanashvily.
\newblock {\em Gravitation}.
\newblock Naukova Dumka, Kiev, 1985.
\newblock In Russian.

\bibitem{CompensationFields}
Dmitri~D. Ivanenko, editor.
\newblock {\em Elementary particles and compensation fields}.
\newblock Mir, Moscow, 1964.
\newblock In Russian.

\bibitem{Konopleva&Popov}
N.~P. Konopleva and V.~N. Popov.
\newblock {\em Gauge fields}.
\newblock Hardwood Academic Publishers, Chur-London-New York, second edition,
  1981.
\newblock (Original Russian edition: Atomizdat, Moscow, 1972 (1~ed.), 1980
  (2~ed.)).

\bibitem{Greub&et_al.-2}
W.~Greub, S.~Halperin, and R.~Vanstone.
\newblock {\em Lie groups, principle bundles, and characteristic classes},
  volume~2 of {\em Connections, Curvature, and Cohomology}.
\newblock Academic Press, New York and London, 1973.

\bibitem{Pham-Mau-Quan}
Pham~Mau Quan.
\newblock {\em Introduction a la g\'eom\'etrie des vari\'et\'es
  diff\'e\-ren\-tiables}.
\newblock Dunod, Paris, 1969.

\bibitem{Slavnov&Fadeev}
A.~A. Slavnov and L.~D. Fadeev.
\newblock {\em Introduction to quantum theory of gauge fields}.
\newblock Nauka, Moscow, 1978.
\newblock In Russian.

\bibitem{Pinl}
M.~J. Pinl.
\newblock Geodesic coordinates and rest systems for general linear connections.
\newblock {\em Duke Mathematical Journal}, 18:557--562, 1951.

\bibitem{ORai}
L.~\'O~Raifeartaigh.
\newblock Fermi coordinates.
\newblock {\em Proceedings of the {Royal} {Irish} {Academy}}, 59
  {Sec.}~A(2):15--24, 1958.

\bibitem{Ivanenko&Sardanashvily-1983}
Dmitri~D. Ivanenko and Gennadi~A. Sardanashvily.
\newblock The gauge treatment of gravity.
\newblock {\em Physics reports}, 94(1):1--45, 1983.

\bibitem{Sardanashvily&Zakharov-1991}
Gennadi~A. Sardanashvily and O.~Zakharov.
\newblock {\em Gauge gravitation theory}.
\newblock World Scientific, Singapore-New Jersey-London-Hong Kong, 1991.

\end{thebibliography}
\bibliographystyle{unsrt}
\addcontentsline{toc}{subsubsection}{This article ends at page}

\end{document}